\documentclass{article}
\usepackage{spconf,amsmath,graphicx}
\usepackage[super]{nth}
\usepackage{tabularx}
\usepackage{subcaption}
\usepackage{booktabs}

\title{Exploring ASR-Based Wav2Vec2 for Automated Speech Disorder Assessment: Insights and Analysis}

\name{Tuan Nguyen$^{1}$, Corinne Fredouille$^{1}$, Alain Ghio$^{2}$, Mathieu Balaguer$^{3}$, Virginie Woisard$^{3,4}$\thanks{This research work was funded by the \textbf{LIAvignon AI Chair}} }
\address{$^{1}$Avignon Université, $^{2}$Aix-Marseille Université, $^{3}$Université de Toulouse, $^{4}$IUC Toulouse} 
\begin{document}
%
\maketitle
\begin{abstract}
With the rise of SSL and ASR technologies, the Wav2Vec2 ASR-based model has been fine-tuned for automated speech disorder quality assessment tasks, yielding impressive results and setting a new baseline for Head and Neck Cancer speech contexts. 
This demonstrates that the ASR dimension from Wav2Vec2 closely aligns with assessment dimensions. 
Despite its effectiveness, this system remains a black box with no clear interpretation of the connection between the model ASR dimension and clinical assessments. 
This paper presents the first analysis of this baseline model for speech quality assessment, focusing on intelligibility and severity tasks. 
We conduct a layer-wise analysis to identify key layers and compare different SSL and ASR Wav2Vec2 models based on pre-trained data. 
Additionally, post-hoc XAI methods, including Canonical Correlation Analysis (CCA) and visualization techniques, are used to track model evolution and visualize embeddings for enhanced interpretability.

\end{abstract}
\section{Introduction}
\label{sec:intro}
In the \nth{21} century, people have a variety of communication choices, which make life easier.
Nonetheless, verbal communication continues to play an irreplaceable role in the culture of humanity.
Because only through verbal communication can we, as humans, fully comprehend and express all the intricate aspects of a subject, including emotions, and more.
The lack of ability to communicate using speech often referred to as a speech disorder, represents a significant loss and necessitates the need for treatment.
Speech disorders can be caused by various reasons such as Parkinson's disease, throat cancer, stroke, etc \cite{desease1,desease2}. 
This leads to various treatment methods specific to different disease stages and causes. 
Moreover, each patient may have different responses or adaptations to the same treatment method.
Regular assessment of speech has to be conducted after a certain period of time to ensure the effectiveness of the treatment method as well as to monitor the patient's condition \cite{desease3}.
However, this assessment process demands substantial resources and expertise. 
Therefore, efforts to develop an automated speech quality assessment architecture as an alternative or support to this process have been increasing in recent years \cite{intro1,intro2}.
Some automatic systems have shown robust performance and stability by learning from expert decisions \cite{sondeslrec,sebastiao}. 

In 2024, Nguyen et al. \cite{tuan} introduced a system that leverages the Automatic Speech Recognition (ASR) based Wav2Vec2 model \cite{facebook}, known for its strong capability in learning speech representations. 
This approach compared self-supervised learning (SSL) and the ASR dimension for speech quality assessment. It is shown that the fine-tuning of SSL models, using the ASR dimension, achieves the best results for the downstream task \cite{tuan}.
Despite its good performance, this assessment system is not perfect, and the actual behavior of the model is not well understood, which could pose significant problems, especially in the medical domain. 
Therefore, it is important to clarify and understand the decision-making process to ensure trustworthiness for humans, making these systems applicable in real-life scenarios.

In this paper, we present the first analysis of the ASR-based Wav2Vec2 model for speech disorder assessment, focusing on the prediction task of both intelligibility and severity scores.
Our study begins with a layer-wise analysis of the model performance to identify which layers are most dedicated to these tasks, providing insights crucial for future research in the community. 
This analysis involves freezing and fine-tuning parts of the model up to selected layers to better understand how training should be approached in future.
Furthermore, we conducted this layer-wise analysis across different versions of Wav2Vec2 models, based on the amount of pre-trained data. 
By comparing their performance in the speech quality assessment tasks, we aim to establish the initial connection between pre-trained SSL data and its impact on speech quality assessment, a question that has yet been addressed in prior work \cite{tuan}. 
This exploration promises benefits not only for the speech disorder community but also for the SSL speech community, offering insights into the effects of data quantity and characteristics on model performance.
In the parallel, we use a post-hoc eXplainable AI (XAI) method to gain more insights.
Specifically, we utilize Canonical Correlation Analysis (CCA) to track how the model evolves across layers.
Finally, we visualize the embedding information to enhance interpretability.


\section{Corpus}
\label{sec:corpus}
This paper utilized four different corpora.
Different variants of Wav2vec2-based ASR were trained using the Common Voice corpus \cite{commonvoice}.
The BREF \cite{bref} corpus was used to develop a phoneme recognition system intended for subsequent layer-wise analysis.
Other analysis experiments of the paper is based on two additional French speech corpora: C2SI \cite{c2si} and SpeeCOmco \cite{SpeeCOmco, speecomco1}, recorded within the context of Head and Neck Cancers (HNC).

\subsection{Common Voice}
\label{subsec:commonvoice}
First introduced in 2019 by Mozilla, Common Voice responded to the problem of training data scarcity for speech technology, which was unavailable for most languages or otherwise prohibitively expensive at that time.
It is a multilingual, open-sourced corpus designed specifically for ASR.
Data collection is conducted through crowd-sourcing, where participants are asked to record their speech by reading sentences displayed on the screen via the project application or website.
In the context of this paper, the French corpus (version 6.1) is used to align with the work of \cite{tuan}.

\subsection{BREF}
\label{subsec:bref}
The BREF corpus, introduced in 1991, is a comparable data for French, similar to other major corpora in different languages such as TIMIT \cite{timit}. 
Specifically designed for assessing automatic speech recognition systems and studying phonological variations, this paper uses the BREF-120 corpus, featuring 120 speakers primarily from the Paris region.
These participants were given a short reading test, which contains sentences collected from LeMonde newspaper.
In total, 115 hours of read-speech data from 65 females and 47 males were collected.

\subsection{C2SI}
\label{subsec:c2si}
C2SI is a French corpus recorded from 2015 to 2017 as part of the Carcinologic Speech Severity Index (C2SI) INCa project,  comprises speech recordings from healthy controls (HC) and patients diagnosed with Head and Neck Cancer. 
Recorded tasks include sustaining /a/ vowels, describing pictures, and reading text passages or pseudowords, facilitating analysis of speech distortion at multiple levels, including phonation, continuous speech production, and prosody-specific aspects.

This paper relies on different sets of recordings from 106 speakers - 82 patients and 24 HC - to conduct experiments. 
The first set is based on passage reading task.
The first paragraph of \textit{La Chèvre de monsieur Seguin}, a tale by Alphonse Daudet, was read by participants.
A group of six experts then listened to these audio recordings and provided individual perceptual evaluations on speech intelligibility and severity.
The evaluation is on a scale from 0 to 10, where a score of 0 represents severe speech disorder or unintelligible speech, and a score of 10 represents normal or highly intelligible speech.
Another set of audio recordings used in this paper was based on the sustained vowel task. 
These recordings contain the production of 3 sustained vowels /a/.
They could provide information on lower dimensions of speech such as voice level, stability, harmonics contents, etc.
\subsection{SpeeCOmco}
\label{subsec:specomco}
Comprising 27 patients suffering from Head and Neck Cancer, Speech and Communication in Oncology (SpeeCOmco) is an additional corpus for C2SI.
Similar to the C2SI passage reading set, participants provided audio recordings of reading \textit{La Chèvre de monsieur Seguin}, which were evaluated by the same panel of experts using the same metrics as in C2SI.
In this study, SpeeCOmco is used to test extended speech quality assessment models, following the approach proposed by \cite{tuan}.

\section{Baseline System}
\label{sec:architecture}
In the study by \cite{tuan}, the authors introduce an architecture using Wav2Vec2-based ASR as the initial component or feature extractor for speech quality assessment tasks.
Subsequently, these features pass through intermediate layers, which include a statistical pooling layer (mean and standard deviation) and two linear layers of size 1024. Finally, a basic linear layer with a dimension of 1 is employed to generate output scores for intelligibility or severity. The model’s performance is measured based on the Mean Squared Error (MSE) between the predicted scores and the ground truth. All layers, including Wav2Vec2, are updated during training to align them with the downstream task space.

They compared the \textit{Wav2Vec2-3K-Large} and \textit{Wav2Vec2-7K-Large} models, both fine-tuned on ASR tasks using the CommonVoice dataset. 
These models were pre-trained on self-supervised tasks using approximately 3000 hours and 7700 hours of healthy speech, respectively. 
The authors observed that starting with Wav2Vec2-based ASR outperformed Wav2Vec2-based SSL in fine-tuning for speech quality assessment, achieving superior results in both intelligibility and severity in the context of HNC patients.
Another finding is also interesting since the 3K model performs better compared with 7K model despite less pre-trained SSL data.
However, it is not totally clear what could caused this difference.

\section{Analytical approaches}
This paper undertakes an analysis of the current baseline of automatic speech quality assessment, as described in the previous section.
Firstly, we add more models using different pre-trained starting points to the original models proposed by Nguyen et al. \cite{tuan}.
Secondly, we extract frame-level features using passage reading , which are then use in Canonical Correlation Analysis (CCA) \cite{ccaoriginal} framework to gain more insights about the models.
This analysis is conducted layer-wise. 
In parallel, different layer embeddings are employed to train for the final task in order to identify the best layer.
Finally, based on the results of CCA and layer-wise training, visualization using scatter plots is performed at the phoneme-level for read speech and at the frame-level for sustained vowels.
\subsection{Additional models for comparative analysis}
\label{subsec:feature_extractor}
To observe the impact of the amount of data used for pre-training SSL, we compared our five different models (comprising two Wav2Vec2-based SSL and three Wav2Vec2-based ASR).

LeBenchmark recently published another pre-trained model, \textit{Wav2Vec2-14K-Large}, which was trained on an additional 7000 hours of data from the 7K-model \cite{lebenchmark2}.
We fine-tuned this new 14K model, along with the previous \textit{Wav2Vec2-1K-Large} pre-trained SSL model, using the Common Voice dataset for the ASR downstream task. 
This process was carried out using the end-to-end approach provided by SpeechBrain \cite{speechbrain}, which is identical to that used by the current baseline.
For a comprehensive comparison, we additionally fine-tuned the 7K model using the Common Voice 6.1 dataset, aligning with other models, instead of following the approach in \cite{tuan} where the available 7K model ASR from SpeechBrain, fine-tuned for Common Voice 14, was used.

Moving forward, we will employ the following labels: \textbf{1K-ASR}, \textbf{3K-ASR}, \textbf{7K-ASR}, and \textbf{14K-ASR} to represent ASR-based models, and \textbf{1K-SSL}, \textbf{3K-SSL}, \textbf{7K-SSL}, and \textbf{14K-SSL} for SSL-based models.

\subsection{Layer-wise training}
\label{subsec:layerwise}
For a more robust investigation of the impact of each layer from the pre-trained Wav2Vec2-based ASR model, we trained this model in layer-wise manner for speech disorder assessment task.
To investigate which layers in the ASR model provide relevant information for downstream tasks, we freeze all layers of the ASR-based Wav2Vec2 model and extract representations layer-wise for downstream training.
In parallel, we conducted partial fine-tuning experiments, we conducted partial fine-tuning experiments, progressing layer by layer from one layer, to two layers, and so forth up to all layers.This approach aimed to explore potential reductions in computational costs, preservation of critical information,etc. 
These findings could provide valuable insights into feature analysis for the target task.

\subsection{Canonical correlation analysis}
\label{subsec:cca}

Canonical Correlation Analysis (CCA) is a statistical method that measure the relationship between continuous-valued vectors by maximizing their linear projections' correlations.
In the context of neural networks, CCA is well-known for evaluating the similarity of representations either between different models or within a single model.
Its ability to remain invariant to linear transformations makes it particularly useful for this purpose.
Consequently, CCA is commonly utilized to investigate the characteristics of deep learning models \cite{cca1,cca2,cca3}.


Since then, multiple variants of CCA have been introduced, but notable ones include Singular Vector CCA (SVCCA) \cite{svcca} and Projection-Weighted CCA (PWCCA) \cite{pwcca}.
Both SVCCA and PWCCA are designed to address the issue where not all dimensions (neurons) of a neural network layer may be utilized or active during the training task.
SVCCA employs singular value decomposition (SVD) to remove low variance neurons that primarily introduce noise. 
On the other hand, PWCCA calculates a weighted mean of the correlation per neuron, assigning higher weights to directions that contribute more to the input. 
Both variants have demonstrated increased robustness compared to the original method.
Given that SVCCA requires determining a threshold for the number of dimensions to be used, we decided to use PWCCA variant instead. 
For simplicity, we will refer to this variant as CCA from this point forward.


CCA is utilized to evaluate the similarity between layer representations of different Wav2Vec2 feature extractors, as described in Section \ref{subsec:feature_extractor},  and their counterparts : (i) from corresponding layer of pre-trained ASR models (\textbf{CCA-ASR}), (ii) pre-trained SSL model (\textbf{CCA-SSL}) and (iii) from the acoustic information of phoneme recognition model (\textbf{CCA-phoneme}).
For this analysis, we exclusively used models with equivalent pre-trained SSL data as the feature extractor variants (1K, 3K, 7K, and 14K).

\subsection{Phoneme encoder}
As observed in previous studies \cite{sondes1,sondesncd,phoneme_relevant1}, phoneme information strongly influences the assessment of severity and intelligibility in speech disorders. 
To compare the information present in the system's feature extractor, we fine-tuned a phoneme recognition model using Wav2Vec2 \textbf{7K-SSL} in an end-to-end approach with the Connectionist Temporal Classification (CTC) loss function.
The model was trained on the BREF corpus, with an 80-10-10\% split between the training, validation, and test sets. It achieved its best performance, with a Phone Error Rate of 3.4\%, on the test set.
This model encodes meaningful phoneme representations, supported by prior research \cite{acoustic, acoustic2}. 
Consequently, the output of the \textbf{last layer of this Wav2Vec2 model (layer 24)} was employed to analyze the phoneme information of the systems using \textit{CCA-phoneme} as described in \ref{subsec:cca}.


\subsection{t-SNE visualization}
Building upon the results obtained from the above analysis methods, we will further examine and \textbf{visualize the representations of last layer of feature extractor Wav2Vec2 - the \nth{24} layer}.
To do that, we applied  t–Stochastic Neighbourhood Embedding (t-SNE) method \cite{tsne} to reduce the information from layer 24 to 2-dimensional plane.
The focus here is to observe the system behavior, in terms of speech representation, at the phoneme level for read speech and at the frame level for the sustained vowel production task.
For the phoneme level, the approach involves averaging all frames of each phoneme utterance using mean and standard deviation to generate a representative vector for the corresponding utterance.

\section{Insights}
All experiments were conducted for both intelligibility and severity assessment targets. 
Due to page limitations, readers should expect similar behavior across both tasks if only a single task is reported without specification.
Additionally, all visualization clustering techniques were applied to the last layer of the feature extractor Wav2Vec2 (layer 24).

For all subsections \ref{subsec:ssldata}, \ref{subsec:ASR-assessment} and \ref{subsec:layerwise_train}, the C2SI reading corpus was used to train the system and compare it with the baseline system. 
This corpus was also utilized to extract embeddings of the system and calculate CCA in subsections \ref{subsec:CCA-ASR}, \ref{subsec:CCA-SSL}, \ref{subsec:phonetic}.
In subsection \ref{subsec:voice}, we used the C2SI corpus sustained vowel audios to extract embeddings and visualize them.
The SpeCOmco corpus was used to report performance in subsections \ref{subsec:ssldata} and \ref{subsec:layerwise_train}, as well as to visualize phoneme information in subsection \ref{subsec:phonetic} with t-SNE.

\subsection{Relationship between pre-trained SSL data and speech quality assessment}
\label{subsec:ssldata}
Following the training and evaluation process outlined in \cite{tuan}, Table \ref{tab:mse} illustrates the performance of additional models detailed in Section \ref{subsec:feature_extractor} using 10-fold validation, compared with the models from the same study.
The results are reported on SpeCOmco corpus using MSE as described in Section \ref{sec:architecture}

Comparing the feature extractors based on SSL, it's evident that the two additional models, 1K-SSL and 14K-SSL, have poorer performance compared to 3K-SSL and 7K-SSL.
It's clear that having thousands of data points less significantly impacts the performance of 1K-SSL on both tasks. 
Despite the additional 7000 hours of data, 14K-SSL fails to perform anywhere close to the levels achieved by 3K-SSL and 7K-SSL.
This result may be due to the additional 7000 hours of data in the 14K-SSL model, which includes Niger-Mali French \cite{lebenchmark2}, leading to a broader range of acoustic information captured by the model. 
This broader scope may not align well with the C2SI corpus, which primarily includes French mainland speakers, making it more challenging for tasks related to comprehensibility or intelligibility.

However, for the severity task, which focuses more on acoustic or low-level speech information, all SSL models have shown similar performance. 
This is logical since SSL models are known for their capability to capture speech representations well, especially acoustic information, resulting in similar performance among them.

\begin{table}[t]
  \caption{MSE results (mean ± std) for severity and intelligibility prediction tasks with different pre-trained models}
  \label{tab:mse}
  \centering
  \begin{tabular}{p{0.35\columnwidth} |p{0.24\columnwidth}| p{0.24\columnwidth}}
    \toprule
    & \textbf{Intelligibility MSE}      & \textbf{Severity MSE}                \\
    \midrule
    \multicolumn{3}{c}{\textit{Feature Extractor Based on Pre-trained SSL}} \\
    3K-SSL \cite{tuan} & 1.65 $\pm$ 0.43 & 2.1 $\pm$ 0.83 \\
    7K-SSL \cite{tuan} & 1.84 $\pm$ 0.49 & 1.83 $\pm$ 0.71 \\
    1K-SSL             & 3.65 $\pm$ 1.44 & 2.30 $\pm$ 0.53 \\
    14K-SSL            & 3.25 $\pm$ 1.4  & 2.23 $\pm$ 0.89 \\
    \midrule
    \multicolumn{3}{c}{\textit{Feature Extractor Based on Pre-trained ASR}} \\
    3K-ASR baseline \cite{tuan} & \textbf{0.73 $\pm$ 0.18} & \textbf{1.15 $\pm$ 0.14} \\
    7K-ASR baseline \cite{tuan} & \textbf{\textit{0.98 $\pm$ 0.26}} & \textbf{\textit{1.15 $\pm$ 0.16}} \\
    1K-ASR             & 0.9 $\pm$ 0.17  & 1.33 $\pm$ 0.21 \\
    7K-ASR             & \textbf{\textit{1.1 $\pm$ 0.23}}  & \textbf{\textit{1.76 $\pm$ 0.50}} \\
    14K-ASR            & 0.86 $\pm$ 0.19 & 1.28 $\pm$ 0.15 \\
    \bottomrule
  \end{tabular}
\end{table}
Looking at the ASR-based models, we observe a similar performance among all ASR models.
Since the ASR models have been fine-tuned with the Common Voice dataset, this may lead to some forgetting of information from the original SSL model, effectively pulling all SSL models towards better alignment with the task.
This suggest that the ASR dimension is closer and more important for speech disorder assessment.
A notable observation is that after ASR fine-tuning, the 7K model exhibits poorer performance compared to the others. 
This could be attributed to more than half of the 7K pre-trained data (4000 hours) leaning towards spontaneous speech, which might be emphasized in ASR task.
\begin{figure*}[t]
  \centering
  \includegraphics[width=.8\linewidth]{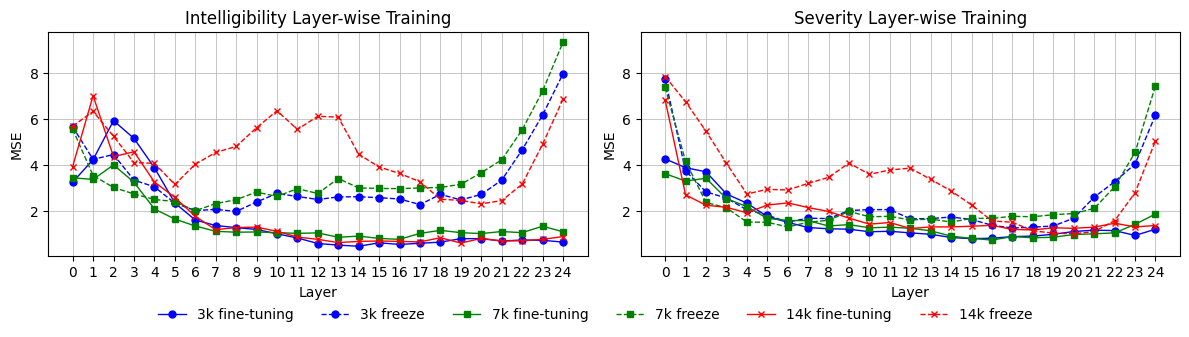}
  \caption{Performance comparison of freeze and fine-tuned layer-wise feature extractor training on speech quality assessment tasks}
  \label{fig:ft}
\end{figure*}
\subsection{Relationship between ASR performance and speech quality assessment}
\label{subsec:ASR-assessment}
Comparing our new 7K-ASR model with the 7K-ASR baseline used in \cite{tuan}, we observe a decline in performance across both tasks, particularly in the severity task where both performance and stability are significantly worse.
This decline can be attributed to the amount of ASR data used to obtain the ASR model. 
As indicated in \cite{tuan}, the pre-trained 7K-ASR model served as a baseline provided SpeechBrain, trained on CommonVoice 14, which includes approximately 400 more hours of ASR data compared to the version used in our study.

On the other hand, the \textit{7K-ASR baseline} model yielded better ASR performance, with a Word Error Rate (WER) of 10.24\%, while our version only achieved a WER of 13.45\%. 
Similar behavior is observed among the 1K, 3K, and 14K ASR models. 
The 3K model achieved the lowest WER, whereas the 1K model had the highest.
Interestingly, this ASR performance pattern aligns with the performance of speech quality assessment, with the 3K model outperforming the others.

The new 7K-ASR model, despite having a better WER than the 1K and 14K models on the same version of CommonVoice corpus (13.45\% WER compared to 16.64\% and 15.52\% WER, respectively), performed worse in the assessment tasks. 
This should be attributed to the amount of pre-trained SSL data, which is more lean towards spontaneous speech, as explained in section \ref{subsec:ssldata}.

\subsection{Layer-wise training analysis}
 \label{subsec:layerwise_train}
Figure \ref{fig:ft} illustrates the results of layer-wise training, both when freezing and fine-tuning the feature extractor model as described in Section \ref{subsec:layerwise}, across different assessment tasks.

There is a similarity in performance between freeze and fine-tuning for the initial layers, indicating that the information in these layers remains relatively stable throughout fine-tuning and could be frozen to faster the process.
Looking at the intelligibility task, there is a notable performance difference between freeze and fine-tuning from higher layers (starting from layer 8).
In contrast, for the severity task, the representations of ASR models at intermediate layers exhibit similar performance with fine-tuning, with some layers achieving identical performance at layer 22.
This suggests that ASR models encapsulate acoustic or low-level information for severity assessment across intermediate layers, requiring only minor adjustments in fine-tuning to achieve convergence.
On the other hand, for intelligibility, the necessary information seems less clear with ASR models, requires fine-tuning to achieve optimal performance. 
Nevertheless, the overall trend for both fine-tuning and freezing indicates that intermediate layers (layer 8 onwards) contain relevant information for speech quality assessment.
However, the 14K-ASR model exhibits a different behavior compared to the others from layer 9 to layer 16.
This unusual behavior in the mid-layers suggests that the additional data from the African accent provide more different speech dimension which is not observed in the other models. 
Further analysis is required which could provide more insight into the impact of data on SSL models and to understand why the ASR performance of the 14K model is inferior to that of the 3K or 7K models, as indicated in \cite{lebenchmark2} despite having more pre-trained data.

\begin{figure}[t]
  \centering
  \begin{subfigure}{\linewidth}
    \centering
    \includegraphics[width=.8\linewidth]{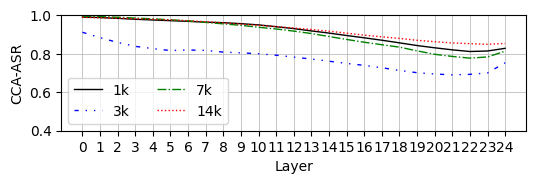}
    \caption{CCA-ASR}
    \label{fig:cca_asr}
  \end{subfigure}
  \begin{subfigure}{\linewidth}
    \centering
    \includegraphics[width=.8\linewidth]{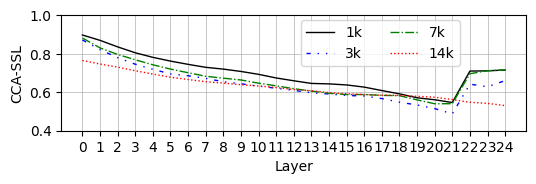}
    \caption{CCA-SSL}
    \label{fig:cca_ssl}
  \end{subfigure}  
  \begin{subfigure}{\linewidth}
    \centering
    \includegraphics[width=.8\linewidth]{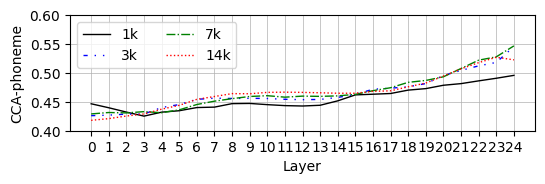}
    \caption{CCA-phoneme}
    \label{fig:cca_phon}
  \end{subfigure}  
  \caption{CCA similarity between fine-tuned feature extractors with pre-trained ASR Wav2Vec2, SSL Wav2vec2 models and phoneme encoder}
\end{figure} 
\subsection{Impact of fine-tuning on the ASR model}
 \label{subsec:CCA-ASR}

As observed in previous sections, the 3K-ASR baseline continues to demonstrate the best performance among all models. Therefore, starting from this section onward, all analyses are conducted using the \textbf{3K-ASR baseline} model.

The CCA-ASR analysis (Fig. \ref{fig:cca_asr}) revealed a consistent and relatively high similarity between the ASR pre-trained model and the feature extractor of the automatic assessment system, gradually decreasing from layer 0 to layer 24 but consistently remained at a minimum level of around 0.7. 
Given this observation, it could be inferred that freezing the upper layers during training is viable, as their similarity is exceptionally high.
This finding provides additional support for the conclusion drawn in \cite{tuan} that pre-trained ASR serves as a strong initialization for both severity and intelligibility assessment.

\subsection{The SSL representation within model}
 \label{subsec:CCA-SSL}
Looking at CCA-SSL (Fig. \ref{fig:cca_ssl}), we can clearly observe a continuance decrease in similarity across layers. 
This aligns with CCA-ASR as well, where similarity primarily relates to ASR. 
However, in the last 3 layers, the CCA score sharply increases to nearly 0.8. 
This is intriguing as \cite{cca2} demonstrated that SSL follows an encoder-decoder style, suggesting that these final layers closely resemble the input as if they are reconstructing the input signal.
Additionally, a notable point is that the 14K model exhibits a different behavior compared to the others, with the CCA score consistently decreasing linearly.

\subsection{Phonetic information in feature extractor}
\label{subsec:phonetic}
When comparing the phoneme information encoded within the feature extractor with the last layer of the phoneme recognition system, we observed a relatively low similarity score of approximately 0.6 across layers (Fig. \ref{fig:cca_phon}).

The data points in Fig. \ref{fig:phone_clustering} represent phoneme utterances, with each point labeled according to phoneme type (consonant or vowel) on the left, and speech quality (severe, mild, and healthy) on the right. 
The plot on the left indicates that feature extractors cannot distinguish between consonants and vowels, corroborating with CCA-phoneme.
However, the plot on the right demonstrates that feature extractors can distinguish between patients based on their speech quality. 
This suggests that unlike ASR, the feature extractors may not clearly distinguish between phonemes (e.g., between vowels and consonants) but may instead capture lower-level phonetic information (nasal, labial, etc), consistent with findings from \cite{sondesncd}.

\subsection{Voice production information in feature extractor}
 \label{subsec:voice}

\begin{figure}[t]
  \centering
  \begin{subfigure}{\linewidth}
    \centering
    \includegraphics[width=.91\linewidth]{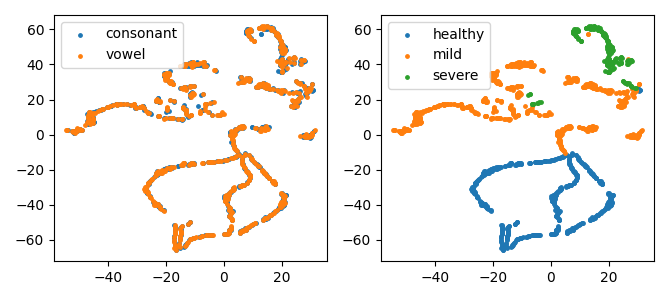}
    \caption{phoneme utterances level}
    \label{fig:phone_clustering}
  \end{subfigure}
  \begin{subfigure}{\linewidth}
    \centering
    \includegraphics[width=.9\linewidth]{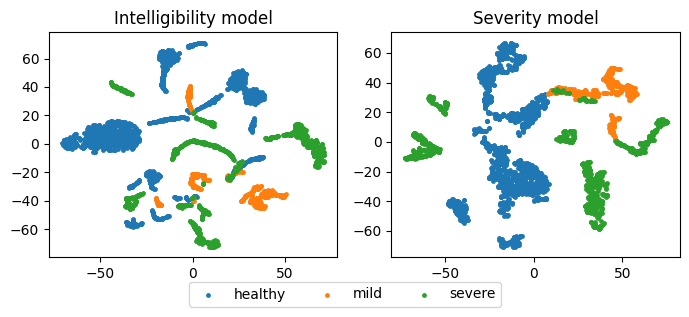}
    \caption{vowel frame level}
    \label{fig:voice}
  \end{subfigure}   
  \caption{2D t-SNE visualization of the last Wav2Vec2 layer}
\end{figure}

Figure \ref{fig:voice} presents the visualization of sustained vowels at the frame level, with each point labeled according to the quality class of the respective patient. 
Since records related to the sustained vowels only contain a single vowel (in this particular case, vowel "a") pronounced continuously, it is typically used to measure voice quality \cite{sustained_vowel}. 
Indeed, this type of audio, especially the stable part of the vowel, makes the measurement of voice characteristics such as jitter and breathiness easier.
As observed in Fig. \ref{fig:voice}, the three levels of speech quality (represented by the three patient groups - healthy, mild, and severe) are distinctly separated, with the severity task showing slightly clearer separation than intelligibility. 
This suggests a strong correlation between voice information and the model decision, which is logical considering that patients suffering from cancer often exhibit significant patterns of fatigue or discontinuous speech whereas healthy speakers do not.
While testing the overall model performance using sustained vowel audio, the model MSE is notably high, with MSE values of 15.16 for the severity task and 17.11 for the intelligibility task. 
This is expected, as the model was trained primarily on read speech. 
Despite Wav2Vec2 ability to capture voice signal details, accurate scoring still requires speech-related dimensions like continuous speech, and phonetic variety.
Combining this with Section \ref{subsec:phonetic}, we can conclude that the model relies not only on speech dimensions such as articulation, resonance, and probably prosody (not studied here), but also needs to incorporate voice information to make final scoring decisions the most accurate.

\section{Conclusion}
This paper presents the first analysis on automatic speech quality assessment using ASR as a pre-trained starting point. 
We found that aligning the domain of pre-trained SSL data with downstream speech tasks (e.g., read speech with read speech) is more critical than the quantity of pre-trained data. 
Additionally, the experiments show a strong correlation between ASR performance and quality assessment, highlighting the impact of not only phonetic features but also low-dimensional voice signals.
Finding of an unusual pattern in the 14K model's layers suggests the need for further investigation into the effects of data quantity on model behavior. 


\newpage
\bibliographystyle{IEEEbib}
\bibliography{refs}

\end{document}